\documentclass[prd,superscriptaddress]{revtex4}
\pdfoutput=1
\usepackage{graphicx}
\ifx\pdfoutput\undefined
\usepackage[dvips,bookmarks]{hyperref}	
\else
\usepackage{hyperref}	
\fi
\begin{document}

\begin{flushright}
{\it IRFU-09-262}\\
{\it SACLAY-T09/178}\\
{\it CERN-PH-TH/2009-215}

\end{flushright}
\title{The Cosmic Ray Lepton Puzzle}
\author{Pierre Brun}\address{CEA, Irfu, Service de Physique des Particules, Centre de Saclay, F-91191 Gif sur Yvette, France}

\author{Gianfranco Bertone}\address{Institut d'Astrophysique de Paris, France. UMR7095-CNRS
UPMC, 98bis Boulevard Arago, 75014 Paris, France \\ \& Institute for Theoretical Physics, University of Zurich, 8057 Zurich, Switzerland}

\author{Marco Cirelli}.\address{Institut de Physique Th\'eorique, CNRS URA2306 \& CEA/Saclay,
F-91191 Gif-sur-Yvette, France \\ \& CERN Theory Division, CERN, Case C01600, CH-1211 Gen\`eve, Switzerland}

\author{Emmanuel Moulin}\address{CEA, Irfu, Service de Physique des Particules, Centre de Saclay, F-91191 Gif sur Yvette, France}

\author{Jean-Fran\c{c}ois Glicenstein}\address{CEA, Irfu, Service de Physique des Particules, Centre de Saclay, F-91191 Gif sur Yvette, France}

\author{Fabio Iocco}\address{
Institut de Physique Th\'eorique, CNRS, URA 2306 \& CEA/Saclay, F-91191
Gif-sur-Yvette, France\\
\& Institut d`Astrophysique de Paris, UMR 7095-CNRS Paris, Universit\'e Pierre
et Marie Curie, boulevard Arago 98bis, 75014, Paris, France}

\author{Lidia Pieri}\address{Department of Astronomy, University of Padua, Vicolo dell'Osservatorio 3, I-35122, Padua, Italy\\  \& INFN Padova - Via Marzolo 8, I-35131, Padua, Italy
}

\begin{abstract}
Recent measurements of cosmic ray electrons and positrons by PAMELA, ATIC, Fermi and HESS have revealed interesting excesses and features in the GeV-TeV range. Many possible explanations have been suggested, invoking one or more nearby primary sources such as pulsars and supernova remnants, or dark matter. Based on the output of the TANGO in PARIS --Testing Astroparticle with the New GeV/TeV Observations in Positrons And electRons : Identifying the Sources-- workshop held in Paris in May 2009, we review here the latest experimental results and we discuss some virtues and drawbacks of the many theoretical interpretations proposed so far.

\end{abstract}

\maketitle

%
\section{Introduction}

Cosmic leptons (electrons and positrons) in the GeV-TeV range represent a useful tool to probe our local Galactic environment. Whatever the sources are (pulsars, Supernova remnants or Dark Matter annihilations), energy losses at these energies are such that the injection spectrum gets severely distorted during the propagation to the Earth, and the contribution of sources at distances larger than about 1 kpc is essentially negligible.

For those reasons, and because the matter/antimatter ratio is higher than for other species such as protons, the leptonic channel has always been thought to be a good way to search for dark matter annihilation debris. In this paper we briefly review recent measurements of cosmic ray leptons as well as some possible interpretations. Note that this brief review cannot be all comprehensive by itself, for more details and discussions we invite the reader to consult the \href{http://irfu.cea.fr/Meetings/TANGOinPARIS}{website} of the TANGO in PARIS workshop.

\section{The measurements}

High energy leptons produce electromagnetic showers when passing through matter, which are observed with the help of calorimeters. In addition, for cosmic ray studies detectors can possess magnetic spectrometers in which case they perform charge identification and separate electrons from positrons, or simple calorimeters in which case they measure the $e^-+e^+$ sum. Recent measurements are displayed in Fig. 1. PAMELA measured the positron fraction ($e^+/ (e^-+e^+)$) between 1 GeV and $\sim$ 100 GeV (R. Sparvoli, \href{http://irfu.cea.fr/Meetings/TANGOinPARIS/slides/sparvoli.pdf}{pdf}, \href{http://irfu.cea.fr/Meetings/TANGOinPARIS/videos/sparvoli.mp4}{mp4}). An unambiguous rise of the fraction is observed above 10 GeV, 
therefore confirming hints found by previous experiments. Other results are obtained without charge identification by the balloon experiment ATIC (J. Isbert, \href{http://irfu.cea.fr/Meetings/TANGOinPARIS/slides/isbert.pdf}{pdf}, \href{http://irfu.cea.fr/Meetings/TANGOinPARIS/videos/isbert.mp4}{mp4}), the Fermi satellite (J. Bregeon, \href{http://irfu.cea.fr/Meetings/TANGOinPARIS/slides/bregeon.pdf}{pdf}, \href{http://irfu.cea.fr/Meetings/TANGOinPARIS/videos/bregeon.mp4}{mp4}) and the ground-based HESS Cherenkov telescopes (K. Egberts, \href{http://irfu.cea.fr/Meetings/TANGOinPARIS/slides/egberts.pdf}{pdf}, \href{http://irfu.cea.fr/Meetings/TANGOinPARIS/videos/egberts.mp4}{mp4}). Data points are shown on the right panel of Fig. 1. The ATIC detector has 22 radiation lengths and a few \% energy resolution, the ATIC collaboration claims the observation of a sharp peaked structure in the $e^-+e^+$ spectrum at $\sim$ 600 GeV. The same measurement as performed by Fermi (8.5 radiation length, $\sim$10\% energy resolution) does not show such a prominent structure, but rather a smooth bump. The interesting region is at the edge of the HESS sensitivity, which observes a break in the spectrum. However, due to $\sim$15\% energy scale uncertainty,  HESS data points hardly constrain the precise shape of the excess. At the moment, the data seem inconsistent. Note however that large systematics (not displayed in Fig. 1) are present. Up to now it is still unclear which spectrum is closer to reality. However, whatever the actual spectrum, these data probably indicate the presence of a nearby yet undetermined cosmic ray leptons source, as exposed below.

	\begin{figure}[t]
	   \centering
	   \includegraphics[height=2.3in]{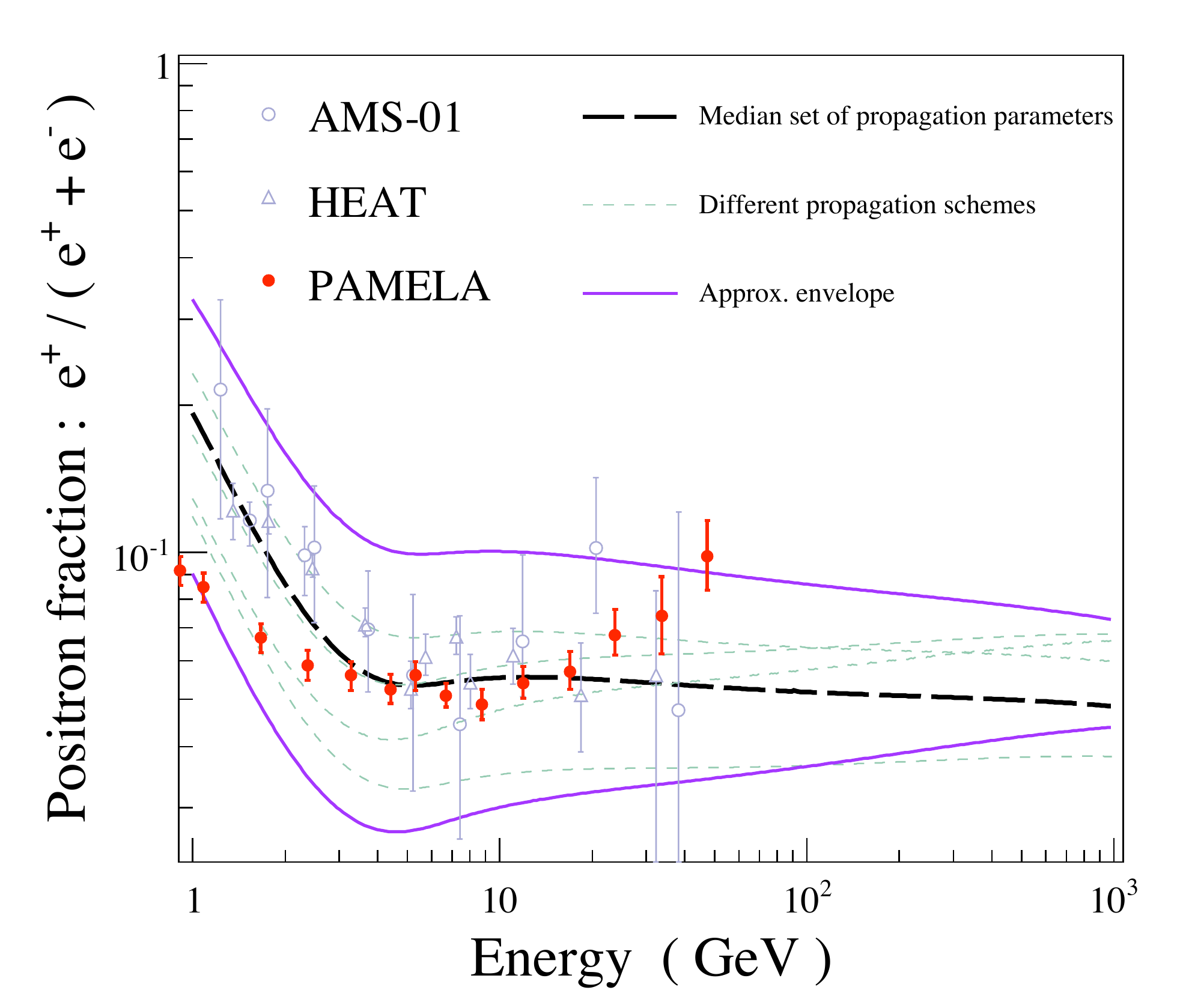}
	   \includegraphics[height=2.3in]{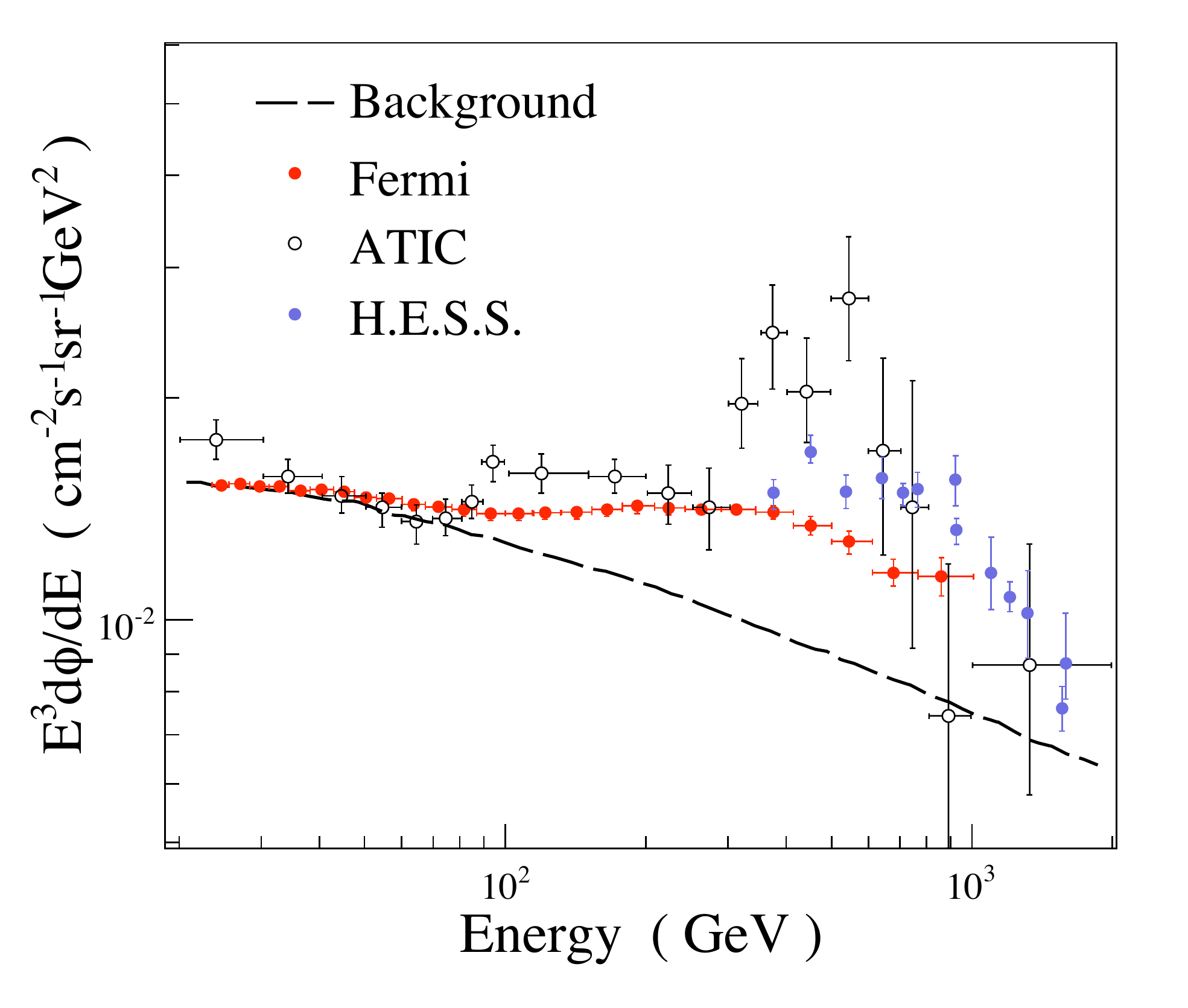}
	      \caption{ Cosmic ray positron fraction (left) and electrons+positrons fluxes (right) (figure from P. Brun \& T. Delahaye, CERN Courier Sep. 2009 issue).}
	       \label{figure_1}
	   \end{figure}

\section{Models for cosmic ray transport}

In the energy range which is considered here (GeV to TeV), it is thought that the bulk of ordinary cosmic rays comes from supernovae and their remnants. These sources are located in the Galactic disk, which is a few tenth of parsecs thick and whose radius is 15 kpc. Once produced and accelerated in the sources, they diffuse inside a flat $\sim$ 10 kpc thick cylinder in which they are deflected by the turbulences of magnetic fields. In addition, they can loose energy and get affected by the convective wind blowing from the disk outwards. The cosmic rays that are directly produced in the sources are called primaries. Some species found in cosmic rays  (such as positrons, antiprotons, boron, etc.) do not originate from stars. These species are produced during the transport of primaries, by spallation reactions onto the interstellar medium.

Two approaches are used to compute cosmic ray fluxes at the Earth in order to compare with the measurements: semi-analytical (D. Maurin, \href{http://irfu.cea.fr/Meetings/TANGOinPARIS/slides/trongmaurin.pdf}{pdf}, \href{http://irfu.cea.fr/Meetings/TANGOinPARIS/videos/strongmaurin.mp4}{mp4}) and fully numerical (A. Strong, \href{http://irfu.cea.fr/Meetings/TANGOinPARIS/slides/strongmaurin.pdf}{pdf}, \href{http://irfu.cea.fr/Meetings/TANGOinPARIS/videos/strongmaurin.mp4}{mp4}). In the first case, mean values are used for the source properties (spatial distribution, intrinsic spectrum, etc.) and propagation parameters (diffusion coefficient, energy loss efficiencies, etc.), and simple boundary conditions are employed. The main advantage of this method is that it is much faster, thus allowing large scans of the propagation parameter space. This essential property permits to compute theoretical uncertainties on the flux as well as on any quantity derived from these parameters. In this framework, it is also easier to perform studies on the spatial origin of sources and local effects. The numerical approach is used in the Galprop code, which is a publicly available tool. In this framework, all type of data is used in a self-consistent way, including 3D gas model, cosmic ray sources, radiation fields, Galactic magnetic field model, etc. This allows in particular the computation of $\gamma$-ray diffuse fluxes, and the production of synchrotron skymaps, which the semi-analytical method still does not do.


This brief description of the computational tools is essential, because when a claim for an excess in some channel is made, those are the methods which are used to determine the expected background. An important point is that whatever the method to estimate the local cosmic ray lepton fluxes, conventional computations cannot account for the observed features in current data. In other words, there is no allowed set of diffusion parameters which reproduces the current leptonic data with secondaries only. Indeed, these estimates lead to the mean secondary flux one can expect at the Earth location, but it is also shown that there should be large fluctuations in the spatial density of cosmic rays. This depends on whether or not a primary source lies in the Earth neighborhood. Ordinary sources such as supernova remnants carry the same baryon number as the whole Universe so that antiparticles should be subdominant. The fact that the positron fraction is $\sim$10\% might be already betraying their nature as secondary species; as most primaries are protons, charge conservation leads to a higher yield of positive charges within secondaries. As electrons and positrons only propagate on short distances of order 1 kpc, the observation of a rise of the positron fraction and a bump in the inclusive spectrum leads to the conclusion that there is a nearby source of high energy $e^+/e^-$ pairs. In the following different possibilities for such a nearby source are reviewed, as well as possible constraints from multi-wavelength and other channels analyses.

\section{Interpretations of the data}

\subsection{Conventional sources}

As explained above, it is suspected that a nearby source of electrons and positrons significantly contributes to the measured flux. We first investigate standard sources which are known to exist, more exotic ideas are discussed in Sec.~\ref{exotic}. As for conventional sources, one can mention two types: electromagnetic sources and non-electromagnetic ones. In the first case, pairs are produced via purely electromagnetic processes. These sources can be $e.g.$ pulsars or $\gamma$-ray binaries. For these sources, counterparts in $\gamma$ rays are expected. The second class of sources imply hadronic processes as well, it can be any type of astrophysical shock, like in supernova remnants. In that case, counterparts are expected in $\gamma$-rays and possibly in the antiproton channel as well.

\begin{figure}[t]
	   \centering
	   \includegraphics[width=0.45\textwidth]{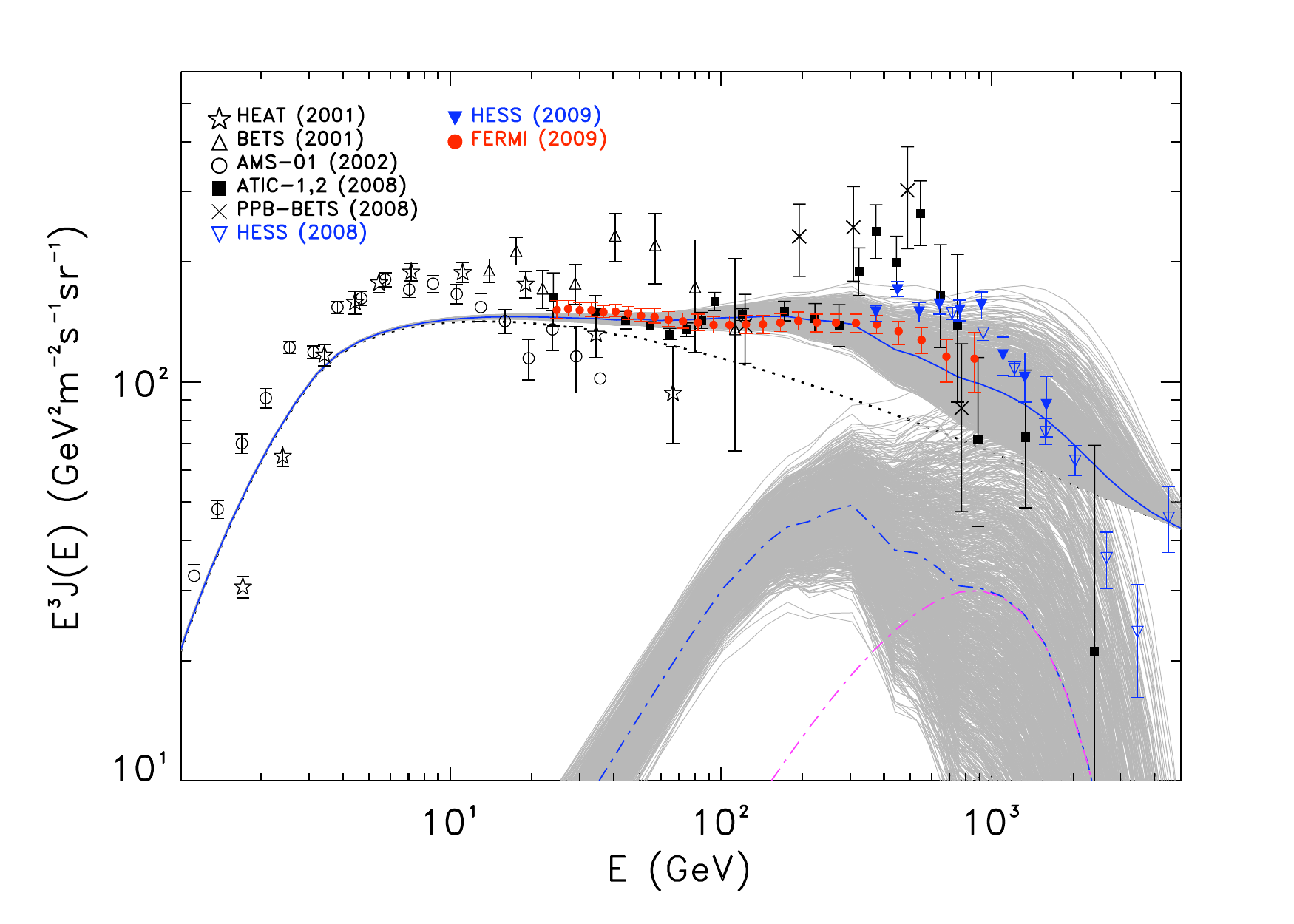}
	   \includegraphics[width=0.45\textwidth]{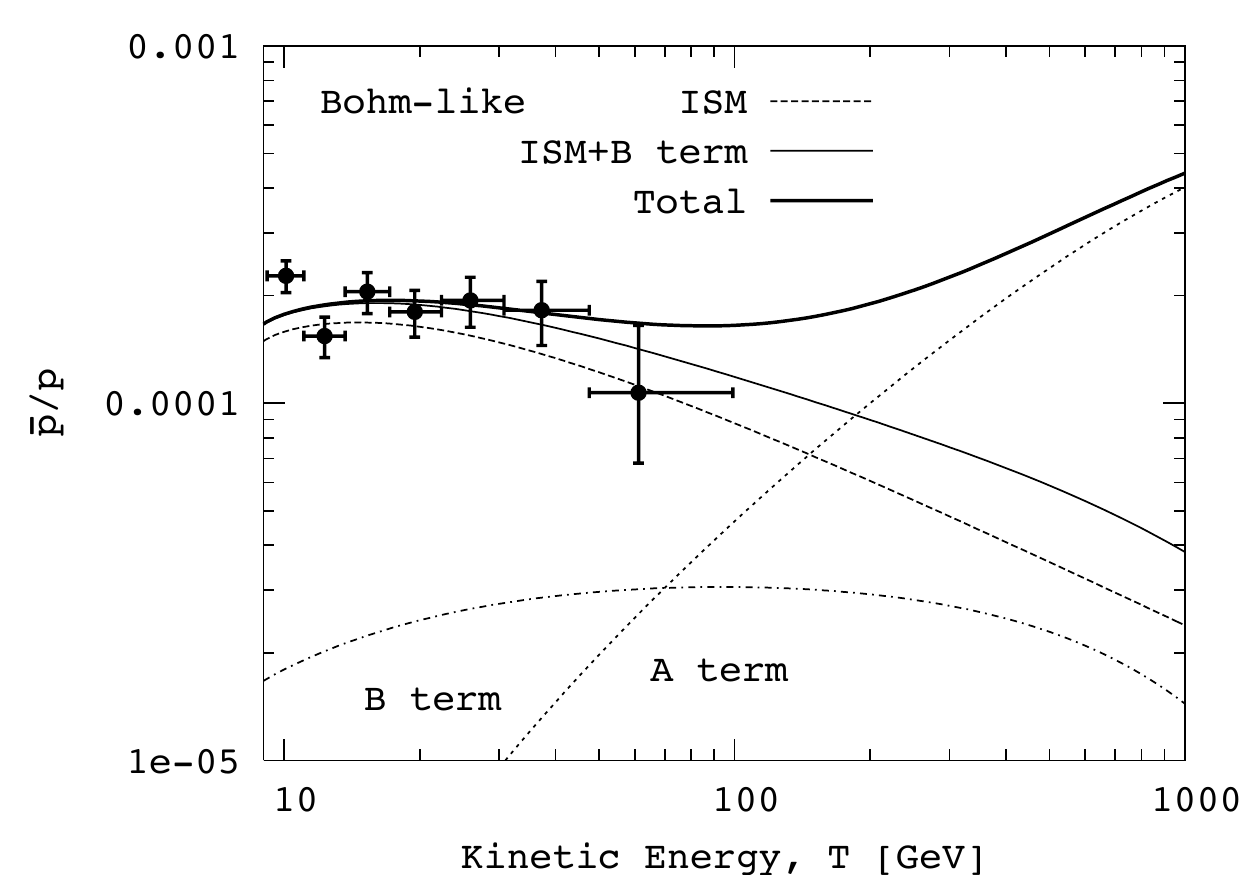}
	      \caption{Fit of the electron data with pulsars (J.Bregeon) and $\bar{p}$ predictions in case of secondary production within a supernovae remnant (P. Blasi).}
	       \label{figure_2}
	   \end{figure}

Pulsars are good candidates for being the nearby source responsible for the excesses. They are rotating and strongly magnetized neutron star, within which $e^\pm$ pairs can be created in magnetic fields or by high energy photons collisions (B. Rudak, \href{http://irfu.cea.fr/Meetings/TANGOinPARIS/slides/rudak.pdf}{pdf}, \href{http://irfu.cea.fr/Meetings/TANGOinPARIS/videos/rudak.mp4}{mp4}). TeV-scale leptons can be produced and accelerated in the environment of the neutron star and released in the interstellar medium provided the matter is diluted enough. For that latter reason, mature pulsars such as Geminga, Loop I, Monogem (and others) are excellent candidates. All the data can be well fitted by adding pulsars to the conventional flux (I. Buesching, \href{http://irfu.cea.fr/Meetings/TANGOinPARIS/slides/bueshing.pdf}{pdf}, \href{http://irfu.cea.fr/Meetings/TANGOinPARIS/videos/bueshing.mp4}{mp4}, see Fig.~\ref{figure_2}). However, there is yet no obvious unique candidate as there are still free parameters in the models. In its first year of data taking, the Fermi satellite discovered plenty new pulsars showing that these objects seem to be ubiquitous in our Galactic environment (D. Smith, \href{http://irfu.cea.fr/Meetings/TANGOinPARIS/slides/smith.pdf}{pdf}, \href{http://irfu.cea.fr/Meetings/TANGOinPARIS/videos/smith.mp4}{mp4}). Unfortunately, a clear counterpart in $\gamma$-ray is not expected as the time scales for photons and charged particles are very different at all levels of the models (production, propagation). 

Concerning supernova remnants, another possibility would be that secondary particles directly produced at the source had  not been previously accounted for properly. In the standard picture, remnant material from the exploding star constitutes a shock in which particles are accelerated. It is now suggested that secondary particles produced in the shock itself could significantly increase the escaping positron fraction (P. Blasi, \href{http://irfu.cea.fr/Meetings/TANGOinPARIS/slides/blasi.pdf}{pdf}, \href{http://irfu.cea.fr/Meetings/TANGOinPARIS/videos/blasi.mp4}{mp4}). Decent fits of the data are obtained within this scenario, which turns out to be falsifiable. This model predicts a rise of the antiproton to proton ratio above 100 GeV. The PAMELA satellite has now precise results up to $\sim$100 GeV and it is foreseen to measure higher energy antiprotons. Therefore it will be possible to test this scenario in the near future. Another conventional interpretation relies on the inhomogeneity of the cosmic ray sources. For instance, it is shown by N.J. Shaviv, E. Nakar \& T. Piran (Phys. Rev. Lett.103:111302, 2009) that the larger concentration of supernova remnants in the Galactic spiral arms and the contribution from a few known supernova remnants can reproduce the measurements.

\subsection{Dark Matter}
\label{exotic}

Although it is definitely possible to reproduce the leptonic data with conventional sources, an extensively studied possibility would be that the features have a dark matter related origin. Observational data show that our Galaxy stars are dipped in a dark halo whose extension is 10 times larger than the luminous disc. The dark matter appears to be non baryonic and could consist of massive yet undiscovered particles. It is believed that these particles were created in the early Universe and now haunt Galaxies. These particles must have very small interactions with conventional matter. It is possible however that they could decay or annihilate. The annihilation or decay product are expected to carry no net Standard Model charge, thus producing standard particles and antiparticles in a 1:1 ratio. These processes happen in the Galactic halo --possibly in the Earth neighborhood--, and the produced particles then diffuse just like conventionally produced cosmic rays, thus enlarging the antiparticle ratio. 

Annihilating dark matter (M.Cirelli, \href{http://irfu.cea.fr/Meetings/TANGOinPARIS/slides/cirelli.pdf}{pdf}, \href{http://irfu.cea.fr/Meetings/TANGOinPARIS/videos/cirelli.mp4}{mp4}) is particularly studied as the involved orders of magnitude nicely converge towards a canonical picture. Dark matter particles appear in Standard Model extensions which are often related to the breaking of the electroweak symmetry. In this framework, Weakly Interacting Massive Particles (WIMPs) have electroweak scale mass which is the same order of magnitude (TeV) as the cosmic lepton features. Primordial self annihilations regulate the cosmological density, thus defining a natural value for the annihilation cross section. As this value is close to the one inferred from electroweak scale interactions, this fact gives credit to the WIMP model. Unfortunately, when computing the related exotic cosmic ray production rate, canonical values give lepton fluxes that are too low by 2-3 orders of magnitude. One needs then to invoke some mechanism for the enhancement of the annihilation rate.  A possibility is an enhancement due to dense dark matter substructures, but this seems somehow unlikely (J. Lavalle, \href{http://irfu.cea.fr/Meetings/TANGOinPARIS/slides/lavalle.pdf}{pdf}, \href{http://irfu.cea.fr/Meetings/TANGOinPARIS/videos/lavalle.mp4}{mp4}). This can be achieved by assuming $e.g.$ a non thermal production in the early Universe (then the annihilation cross section is not constrained from that). A more elegant solution appears for TeV scale WIMPs with the help of the Sommerfeld effect. This happens when masses and couplings have specific values which make the annihilation cross section increase when parameters lead to WIMPs almost-bound states (J. Hisano, \href{http://irfu.cea.fr/Meetings/TANGOinPARIS/slides/hisano.pdf}{pdf}, \href{http://irfu.cea.fr/Meetings/TANGOinPARIS/videos/hisano.mp4}{mp4}). In that case, the colder the dark matter particles, the higher the annihilation cross section. It is essential though to compare exotic cosmic ray leptons with other messengers. For instance, PAMELA precisely measured the antiproton fluxes up to $\sim$100 GeV. Generic dark matter particles annihilations should produce antiprotons as well, so that antiproton measurement can constrain the production rate for exotic cosmic rays. It is shown that the enhancement factor for a conventional WIMP cannot exceed a factor of $\sim$10 not to overproduce antiprotons (F. Donato, \href{http://irfu.cea.fr/Meetings/TANGOinPARIS/slides/donato.pdf}{pdf}, \href{http://irfu.cea.fr/Meetings/TANGOinPARIS/videos/donato.mp4}{mp4}). To save the WIMP interpretation, it is therefore required to assume that hadronic annihilation channels are suppressed, leaving open only leptonic ones. This is the so-called leptophilic dark matter, for which a case can be made from the model-building point of view (either with supersymmetry (N. Fornengo, \href{http://irfu.cea.fr/Meetings/TANGOinPARIS/slides/fornengo.pdf}{pdf}, \href{http://irfu.cea.fr/Meetings/TANGOinPARIS/videos/fornengo.mp4}{mp4}) or within new classes of models (Y. Nomura, \href{http://irfu.cea.fr/Meetings/TANGOinPARIS/slides/nomura.pdf}{pdf})).

Typically, these models postulate the introduction of new particle physics ingredients, building up a richer structure of the dark sector: new force carries that mediate attraction among the dark matter particles (thus producing the Sommerfeld effect), new symmetries that justify the preferential coupling of dark matter to leptons (such as the assignment of a sort of leptonic nature to dark matter itself) or suitable combinations of both.

Decaying dark matter (A.Ibarra, \href{http://irfu.cea.fr/Meetings/TANGOinPARIS/slides/ibarra.pdf}{pdf}, \href{http://irfu.cea.fr/Meetings/TANGOinPARIS/videos/ibarra.mp4}{mp4}) also emerged as an interesting possibility: as long as its half-life is much longer than the age of the Universe, it is conceivable to postulate the dark matter to be unstable, with the tiny fraction decaying today being responsible for the features in the leptonic data. Particle physics models can be somewhat engineered to justify the half-life of about $10^{26}$ seconds that allows to fit the data. Decaying dark matter models have generally the advantage of being less subject to constraints from gamma rays and cosmology such as those discussed below: since signals from decaying dark matter are simply proportional to the density of dark matter particles at the first power (as opposed to its square, as in the case of annihilations), the densest regions of our galactic halo such as the Galactic Center are much less bright and thus constraints based on their observation can often be evaded.

\medskip

	\begin{figure}[t]
	   \begin{center}
	   \includegraphics[width=0.3\textwidth]{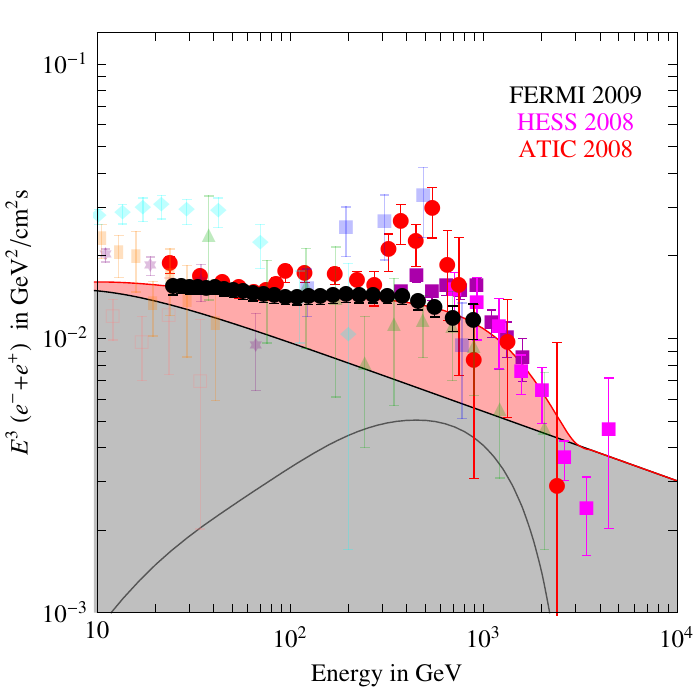}
	   \includegraphics[width=0.45\textwidth]{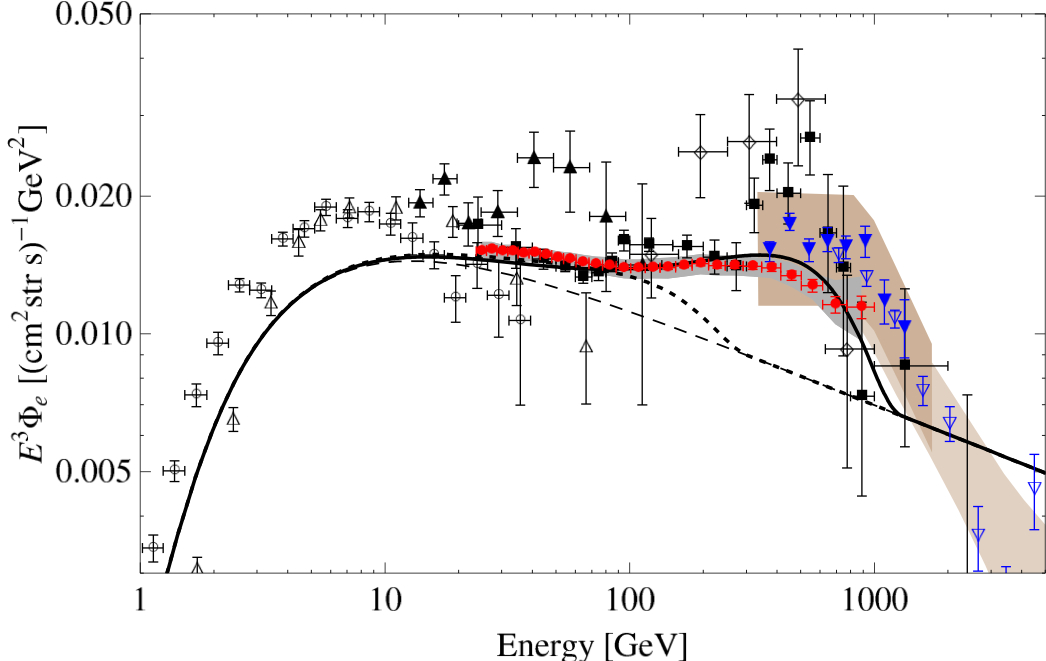}
	   	      \caption{Fit of the electron data with annihilating dark matter (M.Cirelli), decaying dark matter (A.Ibarra).}
	       \label{figure_3}
	       \end{center}
	   \end{figure}

Links with gamma rays of several wavelengths are also under investigation (how the electrons/positrons excesses can be linked to the 511 keV emission at the Galactic center (P. Jean, \href{http://irfu.cea.fr/Meetings/TANGOinPARIS/slides/jean.pdf}{pdf}, \href{http://irfu.cea.fr/Meetings/TANGOinPARIS/videos/jean.mp4}{mp4}) or the WMAP haze (G. Dobler, \href{http://irfu.cea.fr/Meetings/TANGOinPARIS/slides/dobler.pdf}{pdf}, \href{http://irfu.cea.fr/Meetings/TANGOinPARIS/videos/dobler.mp4}{mp4})).

\subsection*{Multi-messenger constraints}

The required large annihilation cross section implies signals in observables other than the charged cosmic rays, and need therefore to be checked against  other astrophysical observations.

Extensive analysis of such multi-messenger constraints in the framework of the latest high-resolution numerical simulations of a Milky Way-like halo have been performed (see e.g. the talk by N. Fornengo).

Prediction of fluxes of i) positrons, ii) anti-protons, iii) gamma-rays from the Galactic Center, iv) gamma-rays from the Galactic halo, v) gamma-rays from the extragalactic halos and subhalos, vi) synchrotron emission due to the propagation of electrons and positrons at the Galactic center are studied.  In particular, the boost factor due to Sommerfeld enhanced cross-section in the low velocity substructures of the Milky Way halo has to be added in a consistent way. The result of such a multi-messenger analysis shows that the typical DM candidates that provide a good fit to the positron data inevitably overproduce antiprotons, gamma-rays or radio emission (or more than one).  

The most conclusive results come from radio and $\gamma$ ray data from the Galactic center and dwarf galaxies. Indeed if these leptons are from dark matter particle collisions, they should be produced in large quantity in other locations. For example, corresponding leptons should be produced at the Galactic center and generate radio waves while interacting with the magnetic field, which are above current observations. In general, even leptophilic dark matter is in conflict with other messengers/wavelength observations, at least within most straightforward models for dark matter distributions and magnetic fields. Some of the new (leptophilic) candidates discussed above can however marginally survive the constraints.To put on even firmer grounds these conclusion, it will be important to better understand the diffuse $\gamma$ ray galactic background at multi-GeV energies (T.Totani, \href{http://irfu.cea.fr/Meetings/TANGOinPARIS/slides/totani.pdf}{pdf}, \href{http://irfu.cea.fr/Meetings/TANGOinPARIS/videos/totani.mp4}{mp4}) thanks to the upcoming observations of the Fermi satellite.

The DM candidates have also to be confronted with constraints coming from cosmological observables: DM annihilations right after the Recombination of hydrogen and the formation of the CMB spectrum, much before structure formation, can modify the ionization state of the  thermal gas at high redshifts. In turn, the presence of additional free electrons modifies the correlation spectra of CMB; the high self-annihilation cross sections needed to explain PAMELA and FERMI/HESS should leave a signature at a level already observable in WMAP5 data. The absence of such signature can put strong constraints in the mass vs cross-section plane, with leptophilic candidates coupling more strongly with the primordial thermal gas and being therefore more disfavored (N. Fornengo, \href{http://irfu.cea.fr/Meetings/TANGOinPARIS/slides/fornengo.pdf}{pdf}, \href{http://irfu.cea.fr/Meetings/TANGOinPARIS/videos/fornengo.mp4}{mp4}).

\section{Outlook}
 
Most likely the lepton excesses observed by PAMELA, ATIC, Fermi and HESS are not caused by conventional secondary cosmic rays, meaning that a nearby source significantly contributes to the local flux. The main classes of candidate are astrophysical ($e.g.$ a nearby pulsar or supernova remnant) or more exotic such as dark matter. The first case seems  favored as one has to invoke non conventional scenarios for dark matter in order to avoid constraints from different observables. 
It is however very exciting to notice that we might be witnessing the direct effects of a TeV cosmic ray source for the first time. Further data analyses of current experiments will certainly provide important information, like whether one can detect a small anisotropy and its direction.  Eventually, future experiments, both on the ground (K. Mannheim, \href{http://irfu.cea.fr/Meetings/TANGOinPARIS/slides/mannheim.pdf}{pdf}, \href{http://irfu.cea.fr/Meetings/TANGOinPARIS/videos/mannheim.mp4}{mp4}) or balloon borne and spatial detectors (L. Derome, \href{http://irfu.cea.fr/Meetings/TANGOinPARIS/slides/derome.pdf}{pdf}, \href{http://irfu.cea.fr/Meetings/TANGOinPARIS/videos/derome.mp4}{mp4}) will certainly offer the opportunity to have a deeper look into this problem.

\begin{acknowledgements}
We would like to warmly thank all the TANGO in PARIS speakers, as well as the members of the organizing committee:  F.~Aharonian, O. de Jager, R. Mohayaee, P. Serpico and J. Silk. We acknowledge the Institut d'Astrophysique de Paris, ANR, IPhT/ CEA Saclay, Irfu/CEA Saclay, GdR Terascale and GdR PCHE for their support.
\end{acknowledgements}

\end{document}